\documentclass[aps,prd,showpacs,preprintnumbers,nofootinbib,twocolumn]{revtex4}

\usepackage{bm}
\usepackage{latexsym}
\usepackage{dcolumn}
\usepackage{amsfonts,amssymb}
\usepackage{graphicx,epsfig}
\usepackage{psfrag}
\usepackage{amsmath,amssymb}

\def\beq{\begin{equation}}
\def\eeq{\end{equation}}
\def\br{\begin{eqnarray}}
\def\er{\end{eqnarray}}
\def\benu{\begin{enumerate}}
\def\eenu{\end{enumerate}}
\def\nn{\nonumber} 
\def\pa{{\partial}}
\def\l{\left}
\def\r{\right}

\def\b  {\beta}

\def\e  {\epsilon}

\def\n  {\nu}
\def\o  {\omega}

\def\p  {\pi}

\def\la {\label}
\def\pa {\partial}
\def\ba{\begin{eqnarray}}
\def\ea{\end{eqnarray}}
\def\f {\frac}

\def\bi {\begin{itemize}}
\def\ei {\end{itemize}}

\def\be {\begin{equation}}
\def\ee {\end{equation}}
\def\la {\label}
\def\be {\begin{equation}}
\def\ee {\end{equation}}
\def\ba {\begin{eqnarray}}
\def\ea {\end{eqnarray}}
\def\bd {\begin{displaymath}}
\def\ed {\end{displaymath}}
\def\nn {\nonumber}
\def\ni {\noindent}
\def\la {\label}

\begin{document}


\title{High frequency quasi-normal modes for black holes with generic 
singularities II: Asymptotically non-flat spacetimes}
\author{Archisman Ghosh}
\email[]{E-mail: archis@iitk.ac.in}
\affiliation{Department of Physics, Indian Institute of Technology Kanpur, \\ Kanpur-208016, India.}
\author{S.~Shankaranarayanan}
\email[]{E-mail: shanki@ictp.trieste.it}
\affiliation{HEP Group, The Abdus Salam International Centre 
for Theoretical Physics,\\ Strada costiera 11, 34100 Trieste, Italy.}
\author{Saurya Das}
\email[]{E-mail: saurya.das@uleth.ca}
\affiliation{Department of Physics, University of Lethbridge,\\
4401 University Drive, Lethbridge, Alberta T1K 3M4, Canada}

\date{\today}


\begin{abstract}
The possibility that the asymptotic quasi-normal mode (QNM)
frequencies can be used to obtain the Bekenstein-Hawking entropy for
the Schwarzschild black hole --- commonly referred to as Hod's
conjecture --- has received considerable attention. To test this
conjecture, using monodromy technique, attempts have been made to
analytically compute the asymptotic frequencies for a large class of
black hole spacetimes. In an earlier work, two of the current authors
computed the high frequency QNMs for scalar perturbations of
$(D+2)$-dimensional spherically symmetric, asymptotically flat, single
horizon spacetimes with generic power-law singularities. In this
work, we extend these results to asymptotically non-flat
spacetimes. Unlike the earlier analyses, we treat asymptotically flat
and de Sitter spacetimes in a unified manner, while the asymptotic
anti-de Sitter spacetimes is considered separately. We obtain master
equations for the asymptotic QNM frequency for all the three cases. We
show that for all the three cases, the real part of the asymptotic QNM
frequency -- in general -- is not proportional to $ln(3)$ thus
indicating that the Hod's conjecture may be restrictive.
\end{abstract}
\maketitle


\section{Introduction} 
\label{sec:introduction}

Classical, damped perturbations about a fixed background which
propagate to spatial infinity are commonly referred to as {\it
quasi-normal modes} (hereafter QNMs) (for excellent reviews, see
Refs. \cite{Nollert:1999,Kokkotas-Schm:1999}). In general, for a
gravitating object like a star, QNM frequencies depend on (i) the
properties of the perturbation such as, the source of the
perturbation, the origin of perturbation, the duration of the
perturbation etc. and (ii) intrinsic properties of the gravitating
object.  However, for the black hole spacetimes, the real (which
corresponds to the frequency of the oscillation) and complex part (which
corresponds to the damping rate) of the QNM frequencies are independent of
the initial perturbations and thereby characterize the black hole
completely. Due to this property, over the last three decades, the
black hole QNM frequencies have attracted a considerable amount of
attention.

Although QNMs are purely classical and have no quantum mechanical
origin, there have been indications -- from two different fronts --
that these carry some information about quantum gravity
\cite{KalyanaRama-Sathi:1999,Horowitz-Hube:1999,Hod:1998,Dreyer:2002}. 
More specifically, it has been shown that QNM can be a useful tool in
understanding the AdS/CFT correspondence. In other words, it has been
shown that there is a one-to-one mapping of the damping time scales
(evaluated via simple QNM techniques) of black holes in Anti-de Sitter
spacetimes and the thermalization time scales of the corresponding
conformal field theory (which are, in general, difficult to compute)
\cite{KalyanaRama-Sathi:1999,Horowitz-Hube:1999}. However, the
primary reason for the recent interest in QNM comes from its
connection to the black hole entropy
\cite{Hod:1998,Dreyer:2002}.

Based on Nollert's numerical result \cite{Nollert:1993}, Hod
conjectured that the real part of the asymptotic QNM frequency should
be treated as the characteristic transition frequency of a
Schwarzschild black hole \cite{Hod:1998}. Using this conjecture and
Bekenstein's conjecture -- the black hole area must be quantized
\cite{Bekenstein:1974a,Bekenstein-Mukh:1995,Kunstatter:2002} -- 
he obtained the Bekenstein-Hawking entropy for the Schwarzschild black
hole. He also showed that this approach is compatible with the
statistical mechanical interpretation of black hole entropy. Later,
Dreyer \cite{Dreyer:2002} reconciled Hod's result with the loop
quantum gravity calculation for the Bekenstein-Hawking entropy (for
criticism of this result, see
Refs. \cite{Domagala-Lewa:2004,Meissner:2004,Dreyer-Mark:2004,Alexandrov:2004}).

Nollert's analytical result was confirmed analytically by two
different methods \cite{Motl:2002,Motl-Neit:2003}. In
Ref. \cite{Motl:2002}, the author used Nollert's continued fraction
expansion for the $4$-dimensional Schwarzschild and showed that the
asymptotic QNM frequencies are given by the following relation:
\beq
\omega_{_{\rm QNM}} = 2 \pi \, i\, T_H \l(n + \frac{1}{2}\r)  + T_H \ln(3)
+ {\cal O}(n^{-1/2}) \, ,
\la{mn01}
\eeq
where $n$ is an integer and $T_H$ is the Hawking temperature of the
Schwarzschild black hole. In Ref. \cite{Motl-Neit:2003}, using the
monodromy technique, the authors confirmed Nollert's result and showed
that Eq. (\ref{mn01}) also holds true for $D$-dimensional
Schwarzschild spacetime. These provided tremendous impetus to verify
Hod's conjecture for a large class of black hole spacetimes. (For a
partial list of references, see
\cite{Birmingham:2003,Birmingham-Carp:2003,Birmingham-Sach:2001,Berti-Cardoso:2004,MaassenvandenBrink:2003,Musiri-Siopsis:2003c,Oppenheim:2003,Cardoso-Kono:2003a,Cardoso-Lemo:2003a,Konoplya:2002,Konoplya:2003,Setare:2003,Konoplya:2004,Setare:2004,Tamaki-Nomu:2004,Suneeta:2003,Paddy:2003,Medved-Martin:2003,Tirth-Paddy:2003,Cardoso-Phd:2004,Kettner-Kuns:2004,Saurya-Shan:2004,Cardoso-Nata:2004,Natario-Schi:2004,Nomura-Tamaki:2005,Daghigh-Kuns:2005,Fernando-Holb:2005,Chakrabarti-Gupt:2005}.)

Hod's conjecture rests heavily on the fact that the real part of the
asymptotic QNM frequencies is proportional to the logarithm of an
integer. The natural question which has lead to a considerable amount
of attention in the field is the following: Is Hod's conjecture
universally valid for all black hole spacetimes?  In an attempt to
address this question, two of the current authors computed high
frequency quasi-normal frequencies for a single-horizon general
spherically symmetric spacetimes with generic singularities and
near-horizon properties
\cite{Saurya-Shan:2004} (hereafter referred to as {\bf I}). For 
these spacetimes, using the monodromy approach, a master equation for
the asymptotic QNM frequency was obtained. It was also shown that the
real part of the high frequency QNM has a logarithmic dependence whose
argument need not necessarily be an integer. However, the result rests
on the assumption of asymptotic flatness. In this work, we extend the
analysis for asymptotically non-flat -- de Sitter and anti-de Sitter
-- spacetimes.

There have been attempts in the literature to obtain high QNM
frequencies for asymptotically non-flat spacetimes
\cite{Cardoso-Nata:2004,Natario-Schi:2004}. Recently, 
Natario and Schiappa \cite{Natario-Schi:2004} have done a detailed
studied of $(D + 2)-$dimensional Schwarzschild de Sitter and Anti-de
Sitter spacetimes. Our treatment differs from that of Natario and
Schiappa's analysis \cite{Natario-Schi:2004} in two ways: (i) As
mentioned earlier, we do not assume any form of the metric except at
the event-horizon and at the origin. At spatial infinity, we assume
that the spacetime is asymptotically flat, de Sitter or Anti-de
Sitter. (ii) Broadly, the numerical results for the asymptotically
flat, de Sitter and Anti-de Sitter spacetimes suggest two classes of
high frequency QNMs (see Sec. (\ref{sec:nr}) for more details). We
demonstrate that the two class of high frequency QNMs can be related
to the two class of boundary conditions (see
Secs. (\ref{sec:mon-bc-cont},\ref{sec:soln}) for more details). Thus,
unlike the earlier analyses, we treat asymptotically flat and de
Sitter spacetimes in a unified manner, while the asymptotic anti-de
Sitter spacetime is considered separately.

The main results of the paper are as follows: (i) We obtain the master
equations for the asymptotic QNM frequency for all the three cases
[see Eqs. (\ref{eq:finres-ds},\ref{eq:finres-afs},
\ref{eq:resultads})]. (ii) We show that for all the three cases, the
real part of the asymptotic QNM frequency, in general, is not
proportional to $ln(n)$ [where $n$ is an integer] thus indicating that
the Hod's conjecture may not be valid for large class of black hole
spacetimes. (iii) We show that, for all the three cases, the high QNM
frequencies have an universal feature. The high QNM frequencies depend
on the parameters of the metric in a specific manner i. e.  $(D q -
2)/2$ [see Sec. (\ref{sec:soln}) for more details]. 

The rest of the paper is organized as follows. In the next section, we
briefly discuss generic properties of the spacetime near the
horizon(s), singularity and spatial infinity. In Sec. (\ref{sec:qnm}),
we briefly discuss the scalar perturbations in the general spherically
symmetric backgrounds and the boundary conditions for the asymptotic
flat, de Sitter and anti-de Sitter spacetimes. In
Sec. (\ref{sec:nrmt}), we discuss the numerical results for the
asymptotically (non-)flat spacetimes and the key properties of the
monodromy technique. In Sec. (\ref{sec:mon-bc-cont}), we discuss the
Stokes lines, contours and boundary conditions for all the three
cases. In Sec. (\ref{sec:soln}), we obtain the asymptotic QNM
frequencies for all the three cases. In Sec. (\ref{sec:specificBH}),
we apply our general results to specific black holes. Finally, we
conclude in Sec. (\ref{sec:disc}) summarizing our results.


\section{Spherically symmetric black hole}
\label{sec:SphSy}

In this section, we briefly review the key properties of spherically
symmetric spacetime. (For more details, we refer the readers to {\bf
I}.)  The line-element for an interval in a $(D+2)$-dimensional
spherically symmetric spacetime $({\cal M}, g)$ (with a boundary $\pa
{\cal M}$) is
\ba
\label{eq:spher-tr}
ds^2 &=& - f(r) dt^2 + \frac{dr^2}{g(r)} + \rho^2(r) d\Omega_D^2 \, ,\\ 
\label{eq:spher-tx}
&=&  f(r)\left[ -dt^2 + dx^2\right] + \rho^2(r) d\Omega_D^2 \, ,
\ea
where $f(r)$, $g(r)$ and $\rho(r)$ are arbitrary (continuous,
differentiable) functions of the radial coordinate $r$, $d\Omega_D^2$
is the metric on the unit $S^D$ and 
\be
x = \int \frac{dr }{\sqrt{f(r) g(r)}} ~,
\label{eq:defx}
\ee
is commonly referred to as tortoise coordinate. The line-element
(\ref{eq:spher-tx}) factorizes into the product of two spaces ${\cal
M}^2 \times S^D$, where ${\cal M}^2$ is the $2$-dimensional spacetime
with Minkowskian topology.

As in {\bf I}, to keep the discussion general, we do not assume any
form for $f(r)$, $g(r)$ and $\rho(r)$. However, we assume the
following generic properties of the spacetime: (i) The spacetime has
a singularity (say, at $r = 0$). Near the singularity, we assume that
the line-element is given by Szekeres-Iyer
\cite{Szekeres-Iyer:1993,Celerier-Szek:2002} metric viz.\
(\ref{eq:gensing1}) below. (ii) The spacetime has one event horizon
(say at, $r = r_h$). Near the event horizon, we assume that the
line-element takes the form of Rindler metric. (iii) Towards the
spatial infinity (say, as $r \to \infty$), we assume that the
spacetime is flat, de Sitter or Anti-de Sitter. In the rest of the
section, we discuss the spacetime properties in these regions.

\subsection{Generic singularity and horizon structure}
\label{sec:gensing}

Near the singularity ($r \to 0$), Szekeres-Iyer
\cite{Szekeres-Iyer:1993,Celerier-Szek:2002} (see also, 
Ref. \cite{Blau-Boru:2004}) had shown that a large class of
spherically symmetric black holes (\ref{eq:spher-tr}) take the
following form:
\ba
\label{eq:gensing1}
ds^2 & \stackrel{r \to 0}{\approx}  & \eta r^{\frac{2p}{q}} dy^2 - 
\frac{4 \eta}{q^2} r^{\frac{2\left(p-q+2\right)}{q}} dr^2 
+ r^2 d\Omega_D^2 \\
&=& \eta x^p \left( dy^2 - dx^2 \right) + x^q d\Omega_D^2 \, , \nn 
\ea
where $\eta = \pm 1, 0$ corresponding to the space, time, null-like
singularities and $p,q$ are the power-law indices which are purely
real. Comparing Eqs. (\ref{eq:spher-tr}, \ref{eq:gensing1}), we have
\beq
\label{eq:fg-r0}
f(r) = - \eta \b^2 r^\frac{2p}{q};~ 
\frac{1}{g(r)} = - \frac{4\eta}{q^2}r^\frac{2(p-q+2)}{q};~
\rho(r) = r \ , ~~
\eeq
and the corresponding tortoise coordinate is 
\be
\label{eq:xsing}
x \sim \int \frac{dr}{r^{1-\frac{2}{q}}} = r^{\frac{2}{q}} \, .
\ee
It is interesting to note that $x$ depends only on $q$ and not on $p$.

Near the horizon, we assume that the general spherically symmetric
line-element (\ref{eq:spher-tr}) takes the form of Rindler metric:
\beq
ds^2 \to - \kappa_h^2 \gamma^2 dt^2 + d\gamma^2 + 
\rho^2(r_h) \, d\Omega^2_D \, ,
\label{eq:evehor}
\eeq
where 
\beq
\label{eq:sur-grav}
\gamma = \frac{1}{\kappa_h} \sqrt{f}\, , 
\, \frac{d\gamma}{dr} = \frac{1}{2\kappa_h} \frac{d_r f}{\sqrt{f}} \, ,  \, 
\kappa_h = \frac{1}{2} \left(\sqrt{\frac{g(r)}{f(r)}} d_r f\right)_{r_{h}} \, ,
\eeq
and the tortoise coordinate is given by 
\be
\label{eq:xhor}
x \sim \frac{1}{2 \kappa_h} \ln \left( r - r_{h} \right) \, .
\ee
%

\subsection{Spatial Infinity}
\label{sec:spatinf}

In this sub-section, we discuss spatial asymptotic properties for the
three cases -- asymptotically flat, de Sitter and Anti-de Sitter
spacetimes. \\

\underline{\it Asymptotic flat spacetimes:} \\
 
A spacetime $({\cal M}, g)$ is said to be asymptotically flat if it is
asymptotically empty, i. e. $R_{\mu\nu} = 0$ in an open neighborhood
of $\pa {\cal M}$ in ${\cal M}$.  A static observer in these
spacetimes is bounded by the past/future event horizons and ${\cal
I}^{\pm}$.

The line-element (\ref{eq:spher-tr}) towards the spatial infinity
takes the following form:
\beq
\label{eq:Asyfla}
ds^2 \simeq -  dt^2 + dr^2 + r^2 d\Omega_D^2 \, .
\eeq
Comparing Eqs. (\ref{eq:Asyfla}, \ref{eq:spher-tr}), we have
\be
\label{eq:fgaf}
f(r) = g(r) \sim 1; \, \rho(r) \sim r; \, x \sim r \, .
\ee
For these spacetimes, from our assumption, there exists only one
physical (real, positive) horizon at $r=r_h$ whose surface gravity is
given by the relation (\ref{eq:sur-grav}). \\

\underline{\it Asymptotic de Sitter spacetimes:} \\

A static observer is bounded by the past/future event horizons and
past/future cosmological horizons. Although the coordinate $r$ goes up
to $\infty$, the physical region terminates at the cosmological
horizon ($r=r_c$, $x \to \infty$). For computation of conserved
charges in asymptotically de Sitter spacetimes, see for instance,
Ref. \cite{Balasubramanian-deBo:2001a}.

The line-element (\ref{eq:spher-tr}) at spatial infinity takes the
following form:
\beq
\label{eq:AsydS}
ds^2 \simeq  \frac{r^2}{\ell^2} dt^2 - \frac{\ell^2}{r^2} dr^2
+ r^2 d\Omega_D^2 \, ,
\eeq
where $\ell^2$ is related to the $(D + 2)$-dimensional {\it positive}
cosmological constant, i. e.,
\beq
\Lambda = \frac{D (D + 1)}{2 \ell^2} \, .
\eeq
Comparing Eqs. (\ref{eq:spher-tr},\ref{eq:AsydS}), we have
\be
\label{eq:fgdS}
f(r) = g(r) \sim - \frac{r^2}{\ell^2}; \, \rho(r) \sim r; \, 
 x \sim x_0 + \frac{\ell^2}{r} \, .
\ee
In this case, unlike the previous two cases, the spacetime contains
two physical horizons -- event horizon ($r=r_h$) and cosmological
horizon (say, at $r=r_c$). The surface gravity at $r_h$ is given by
(\ref{eq:sur-grav}). The surface gravity of the cosmological horizon 
is 
\be
\label{eq:defkc}
\kappa_c \equiv \frac{1}{2} \left( \sqrt{\frac{g(r)}{f(r)}} 
\frac{df(r)}{dr}\right)_{r=r_c} \, ,
\ee
which we set to be negative.  \\

\underline{\it Asymptotic Anti-de Sitter spacetimes:} \\

A static observer in these spacetimes is bounded by the past/future
event horizons and the finite spatial boundary. For computation of
conserved charges in asymptotically anti-de Sitter spacetimes, see 
Ref. \cite{Ashtekar-Magn:1984,Ashtekar-Das:1999}.

The line-element (\ref{eq:spher-tr}) near the spatial infinity takes the
following form:
\beq
\label{eq:AsyAdS}
ds^2 \simeq  - \frac{r^2}{\ell^2} dt^2 + \frac{\ell^2}{r^2} dr^2
+ r^2 d\Omega_D^2 \, ,
\eeq
where $\ell^2$ is related to the $(D + 2)$-dimensional {\it negative}
cosmological constant, i. e.,
\beq
\Lambda = - \frac{D (D + 1)}{2 \ell^2} \, .
\eeq
Comparing Eqs. (\ref{eq:spher-tr},\ref{eq:AsyAdS}), we have
\be
\label{eq:fgAdS}
f(r) = g(r) \sim \frac{r^2}{\ell^2}; \, \rho(r) \sim r; \, 
x \sim x_0 - \frac{\ell^2}{r} \, ,
\ee
where $x_0$ is asymptotic value of $x$ and, in general, depends on the
negative cosmological constant and black hole properties. For these
spacetimes, as-well, there exists only one physical horizon, at
$r=r_h$ with surface gravity $\kappa_h$ as in (\ref{eq:sur-grav}). \\

\section{Quasi-normal modes}
\label{sec:qnm}

In this section, we obtain the differential equation corresponding to
scalar field propagating in $(D + 2)-$dimensional spherically
symmetric spacetime (\ref{eq:spher-tr}) and discuss the ``canonical''
boundary conditions corresponding to the three cases.  [By canonical,
we mean the boundary conditions applied in the real $x$ i.e. in the
range $(-\infty,\infty)$.]

\subsection{Scalar perturbations}
\label{sec:scalarpert}

The perturbations of a $(D + 2)$-dimensional static black holes
(\ref{eq:spher-tr}) can result in three kinds -- scalar, vector and
tensor -- of gravitational perturbations (see for example,
Ref. \cite{Kodama-Ishi:2003b}). The higher dimensional tensor
perturbations, which is of our interest in this work, correspond to
the well-known four-dimensional Regge-Wheeler potential\footnote{Note
that we loosely refer to these perturbations as scalar. This is due to
the fact that there is a one-to-one correspondence between the
massless scalar field propagating in the fixed background and the
tensor perturbation equations derived from the linear perturbation
theory.}.  The evolution equation for these perturbations correspond
to the equation of motion of the massless, minimally coupled scalar
field, i. e.,
\be
\label{eq:eq0}
\Box \Phi \equiv \frac{1}{\sqrt{-g}}~\partial_{\mu} 
( \sqrt{-g}g^{\mu\nu}\partial_{\nu} \Phi ) =  0 \, .
\ee
The symmetry of the line-element (\ref{eq:spher-tr}) allows us
to decompose the scalar field modes as:
\be
\label{eq:subst}
\Phi(x^{\mu}) = \rho(r)^{-\frac{D}{2}} \, R(r) \, e^{i \omega t} \, 
Y_{lm_1 ... m_{D - 1}} \, ,
\ee
where $Y_{lm_1..m_{D-1}}$ are Hyper-spherical harmonics and the
function $R(r)$ satisfies the differential equation
\be
\label{eq:rweqn}
\frac{d^2 R(r)}{dx^2} + \left[ \omega^2 - V(r) \right] R(r) = 0 \, ,
\ee
where $x$ is given by (\ref{eq:defx}), $r$ is understood to be $r(x)$
and
\ba
\label{eq:rwpot}
V(r) &=& \frac{l(l+D-1)}{\rho^2(r)}~f(r) + 
\left( \frac{D}{2} \right) \rho(r)^{-\frac{D}{2}} \sqrt{f(r) \,g(r)} \nn \\
&\times& \frac{d}{dr} \left\{ \rho(r)^{\frac{D-2}{2}} \, 
\frac{d\rho(r)}{dr} \sqrt{f(r) \, g(r)} \right\} \, ,
\ea
is the generalized Regge-Wheeler potential.  Before discussing the
boundary conditions, it is important to know the structure of the
singularities of the differential equation at the three -- $0, r_h,
r_c ({\rm or~}\infty)$ -- points. For the differential equation
(\ref{eq:rweqn}), we have: (i) $r = r_h$ is a regular singular
point. (ii) $r \to \infty$ (relevant for the asymptotically flat and
Anti-de Sitter spacetimes) is an irregular singular point.  $r = r_c$
is a regular singular point. (iii) In order for $r = 0$ to be a
regular singular point it can be easily shown that $p, q$ {\it must}
satisfy the following conditions (see {\bf I}):
\beq
q>0 ~~\mbox{and}~~p - q + 2 > 0   \, .
\label{eq:pq-cond} 
\eeq
%

\subsection{Canonical boundary conditions}
\label{sec:bc}

QNMs are solutions to the differential equation (\ref{eq:rweqn})
subject to a specific (physically motivated) boundary
conditions\footnote{Note that, we choose the sign of the exponent
$e^{i \omega t}$ in equation (\ref{eq:subst}) above, we fix the sign
$\Im(\omega)>0$. This is because $\Im(\omega)<0$ leads to solutions
growing with time, which are unphysical.}. The Wronskian of these
modes vanish which gives the corresponding QNM frequencies
\cite{Kokkotas-Schm:1999}.

In order to obtain the QNM frequencies corresponding to the
differential equation (\ref{eq:rweqn}), we need to know the mode
functions. The Regge-Wheeler potential is a complicated function of
$f(r), g(r)$ and $\rho(r)$. Hence, in general, the differential
equation (\ref{eq:rweqn}) can not be solved exactly. In such a
situation, asymptotic analysis is a useful tool to extract some
physical information. In the case of QNM, the asymptotic analysis
also provides us with the identification of the boundary conditions.\\

\underline{\it Asymptotic flat spacetimes:} \\  

The generalized Regge-Wheeler potential decays exponentially near the
event-horizon and as a power-law near spatial infinity, i.e.,
\beq
\label{eq:RW-afshor}
V[r(x)] \stackrel{x \to -\infty}{\simeq}  \exp\l(2 \kappa_h x\r)\, ; 
\, V[r(x)] \stackrel{x \to \infty}{\simeq} \frac{1}{x^2} \, .
\eeq
Thus, the general solution to Eq. (\ref{eq:rweqn}) near the two
boundary points is a superposition of plane-waves:
\beq
R[x] \stackrel{x \to \pm\infty}{\sim} 
C_1^{\pm} \exp(i \omega x) + C_2^{\pm} \exp(-i \omega x) \, ,
\label{eq:plawav1}
\eeq
where $C_1^{\pm}, C_2^{\pm}$ are the constants determined by the
choice of the boundary conditions. QNM boundary condition corresponds
to 
\beq
C_2^{-} = 0 ; C_1^{+} = 0  \, 
\Longrightarrow  \, 
R(x)\sim e^{\pm i\omega x} \, \, \, {\rm as} \, \, \, 
x\rightarrow \mp \infty\, .
\label{eq:bc1}
\eeq
Physically, the boundary conditions mean that no classical radiation
emerge from the (future) event horizon, and no radiation originates at
spatial infinity. \\

\underline{\it Asymptotic de Sitter spacetimes:} \\  

The generalized Regge-Wheeler potential decays exponentially near the
two -- event and cosmological -- horizons i. e.,
\beq
\label{eq:RW-dshor}
V[r(x)] \stackrel{x \to -\infty}{\simeq}  \exp\l(2 \kappa_h x\r)\, ;
\, V[r(x)] \stackrel{x \to \infty}{\simeq} \exp\l(-2 |\kappa_c| x\r) \, .
\eeq
As in the case of asymptotically flat spacetime, the solution near the
two boundary points is a superposition of plane-waves
(\ref{eq:plawav1}) and hence, the boundary conditions are same as that
of asymptotically flat space (\ref{eq:bc1}).

Physically, the boundary conditions mean that no classical radiation
emerge from the event and cosmological horizons. \\

\underline{\it Asymptotic anti-de Sitter spacetimes:} \\
   
The generalized Regge-Wheeler potential decays exponentially in the
event horizon, however the potential grows at spatial infinity:
\br
\label{eq:RW-ads1}
V[r(x)] \stackrel{x \to -\infty}{\simeq}  \exp\l(2 \kappa_h x\r)\, ;
\, V[r(x)] \stackrel{x \to x_0}{\simeq} \frac{j_{\infty}^2-1}{4 (x - x_0)^2} 
\, , 
\, .
\er
where at spatial infinity $x$ goes as
\beq
\label{eq:xasympads}
x \sim x_0 - \frac{\ell^2}{r} \quad \mbox{and} 
\quad j_{\infty} = D + 1 \, .
\eeq
[Even though, this is a standard result and can be found in other
references (see, for instance, Ref. \cite{Winstanley:2001}), we have
given the relevant steps in Appendix \!(\ref{app:Sol-AdS}) for
completeness.] Thus, the general solution to Eq. (\ref{eq:rweqn}) near
the two boundary points is given by
\br
\!\!\!R[r(x)] & \stackrel{x \to -\infty}{\simeq}&
C_1^{-} \exp(i \omega x) + C_2^{-} \exp(-i \omega x) \, \, \, \, \,  \\
&\stackrel{x \to x_0}{\simeq}&
C_1^{+} (x_0-x)^{-\frac{D}{2}} + C_2^{+} (x_0-x)^{\frac{D}{2}+1} \, . 
\, \, \, \, \, 
\label{eq:plawav2}
\er
QNM boundary condition corresponds to
\begin{subequations}
\br
\label{eq:bc2a}
C_2^{-} = 0   & \Longrightarrow & 
R(x)\sim e^{ i\omega x}\quad {\rm as} \quad x\rightarrow - \infty\,  \\ 
C_1^{+} = 0 &\Longrightarrow & 
R(x) \simeq 0 \quad {\rm as} \quad x\rightarrow x_0 .
\label{eq:bc2}
\er
\end{subequations}
Physically, the boundary conditions mean that no classical radiation
emerge from the (future) event horizon, and modes reflect at the
spatial boundary. It is necessary to have a reflecting boundary
conditions for the following reasons: (i) The perturbation equation
(\ref{eq:rweqn}) is obtained from the first order perturbation theory
implying that the stress-tensor of the perturbation is small compared
to the background. If $C_1^{+} \neq 0$, then this assumption is
violated and leads to inconsistency. (ii) To the linear order, the
perturbations conserve energy-momentum. The exponential growth of the
modes would violate energy conservation \cite{Avis-Isha:1978}.

\section{Numerical results and monodromy technique}
\label{sec:nrmt}

In this section, we briefly discuss the numerical results of the
asymptotic QNM frequencies. We also briefly discuss the monodromy
technique which has proven to be useful to analytically calculate
asymptotic QNM frequencies.

\subsection{Numerical results}
\label{sec:nr}

As mentioned earlier, the perturbation equation (\ref{eq:rweqn})
cannot be solved exactly and one has to resort to approximation
methods to obtain analytical results for QNM frequencies. Broadly, the
analytical/numerical approaches in obtaining the QNM frequencies can
be classified into four categories: (i) Approximating the
Regge-Wheeler potential with some simple functions to obtain the exact
QNM frequencies. (ii) Solving the perturbation equation iteratively by
using the well-known techniques like WKB or Born approximation (iii)
Continued fraction technique. (iv) Monodromy technique. (For an
excellent review of the above techniques, see
Ref. \cite{Nollert:1999}.) It is needless to say that nearly all of
these approaches have their own limitations; certain analytical
techniques are useful in certain QNM frequency range while certain
others techniques for certain other ranges. For instance, the
monodromy technique -- which is of our interest in this work -- has
proven to be useful for obtaining asymptotic QNM frequencies.

The numerical results for asymptotic QNM frequencies have been obtained
by various authors following Nollert's seminal result
\cite{Nollert:1993} for $4$-dimensional Schwarzschild. 
Nollert's results have been extended to other dimensions by Cardoso
and his collaborators \cite{Cardoso-Lemo:2003a,Cardoso-Lemo:2003c}
(see also Ref. \cite{Xue-Shen:2003}). In the case of asymptotic AdS
spacetimes, the first numerical calculation was done by Horowitz and
Hubeny \cite{Horowitz-Hube:1999} for $4,5$ and $7$-dimensional
Schwarzschild-AdS black holes in the large black hole limit.  Their
results have been extended by host of other authors for large,
intermediate and small black hole limits
\cite{Cardoso-Lemo:2001a,Konoplya:2002,Cardoso-Kono:2003a,Berti-Kokka:2003b}. 
For the asymptotic dS spacetimes, the numerical results are obtained
for Schwarzschild-de Sitter spacetime by various authors
\cite{Cardoso-Lemo:2003d,Yoshida-Futa:2003,Zhidenko:2003,MaassenvandenBrink:2003b}.

Broadly, two classes have emerged from these numerical results:
\begin{enumerate}
\item For the 
asymptotically flat and de Sitter spacetimes, the numerical results
for the high-frequency QNMs indicate that
\beq
\Im(\omega_{_{QNM}}) \gg \Re(\omega_{_{QNM}}) \, .
\label{eq:FrCon1}
\eeq
\item For the asymptotically Anti-de Sitter spacetimes, the numerical 
results for the high-frequency QNMs indicate that
\beq
\Im(\omega_{_{QNM}}) \sim \Re(\omega_{_{QNM}}) \, .
\label{eq:FrCon2}
\eeq
\end{enumerate}

In the previous section, we showed that the QNM boundary conditions
for the asymptotic flat and de Sitter spacetimes are identical,
however, it is different in the case of asymptotic Anti-de Sitter
spacetimes.  Using the numerical results and the boundary condition,
it is easy to note the following: The structure of the asymptotic QNM
frequencies crucially depend on the choice of the boundary conditions.
 
In Sec. (\ref{sec:soln}), we obtain the high QNM frequencies for the
three cases. Unlike the earlier analyses, we treat asymptotically flat
and de Sitter spacetimes in a unified manner, while the
asymptotically anti-de Sitter spacetime is considered separately.

\subsection{Monodromy technique}
\label{sec:monodromy}

QNMs are damped modes whose frequencies are complex. This implies that
the QNM frequencies will have positive imaginary parts. It follows
that each QNM eigenfunction $\exp(i \omega_n t) R[r(x)]$ (for
instance, in the asymptotically flat spacetime) will grow/decay
exponentially both towards infinity and at the horizon. Thus, as the
QNM mode traces from the horizon to infinity the modes can grow
exponentially within a small region. In other words, the modes which
are exponentially suppressed in a region can grow exponentially in the
nearby region\footnote{A curve which separates these two regions is
referred to as ramification line and the two regions separated by the
ramification line are called ramification regions. In the language of
complex analysis, these ramification lines are nothing but the branch
cuts.}. This implies that analytical/numerical techniques require
exponential precision. Monodromy technique has proved to be a powerful
and flexible approach in obtaining the high frequency QNMs.

Monodromy technique has five key steps: 
\begin{enumerate}
\item  {\it Analytically continue the QNMs in the 
complex $r$ (or $x$) plane.}

This allows one to study the properties of these modes near the
singularity ($r = 0$). In the monodromy technique, unlike other
techniques, both $\omega$ and $x$ are complex.
\item {\it Map the QNMs from the real $x$ to the complex 
$\omega x$ plane. }

Mathematically, this involves finding QNM solutions to the following
differential equation
\be
\label{eq:rxweqn}
\frac{d^2 R(r)}{d (\omega x)^2} + \left[1 - 
\frac{V(r)}{\omega^2} \right] R(r) = 0 \, .
\ee
Even though, Eqs. (\ref{eq:rweqn}, \ref{eq:rxweqn}) look identical,
operationally they are quite different: Firstly, $\omega$ is a complex
number and hence the independent variable $\omega x$ is complex even
if $x$ is purely real. Secondly, since the independent variable
($\omega x$) is complex, the issues of the existence, uniqueness of
solutions [to the differential equation (\ref{eq:rxweqn})] satisfying
the boundary conditions is non-trivial compared to that of
Eq. (\ref{eq:rweqn}).

In the large asymptotic limit of $|\omega|$, Eq. (\ref{eq:rxweqn}) 
can be approximated to  
{\small
\beq
\frac{d^2 R(r)}{d (\omega x)^2}  = - R(r) ~~ 
\Longrightarrow ~~ R[\omega x] \sim \exp(\pm i \omega x) \, .
\label{eq:Asym-Mono2} 
\eeq
}

\noindent Thus, in the high-frequency limit, QNMs can be approximated to be the
superposition of plane-waves in whole of complex $\omega x$ plane
except at the isolated singularities or branch cuts .
 
Setting $\omega = \omega_R + i \omega_I, x = x_R + i x_I$, the
asymptotic modes (\ref{eq:plawav1}) take the following form:
\br
R[\omega x] &\sim & \exp\l[\pm \, i  
(\omega_R x_R - \omega_I x_I ) \r] \nn \\
&\times& \exp \l[\mp (\omega_R x_I + \omega_I x_R) \r] \, .
\label{eq:Asym-Mono3}
\er
%
\item {\it Identify the Stokes line, contours in the $\omega x$ 
complex plane.}

The Stokes lines are defined by the condition $\Im(\omega x) = 0$
\footnote{The Stokes lines are multi-valued near the
horizon where the function $x$ is multi-valued.}.  Under the condition,
Eq. (\ref{eq:Asym-Mono3}) gets simplified to
\beq
R[\omega x] \sim \exp\l[\pm  \, i(\omega_R x_R - \omega_I x_I ) \r] 
\label{eq:Asym-Mono4}
\eeq
Thus, all along the Stokes line the modes are oscillating without any
exponentially growing/decaying solutions. In the next section, we will
show that even near the singularity the solutions to
Eq. (\ref{eq:rxweqn}) are plane waves.
\item {\it Use the numerical results for the 
high-frequency QNMs to translate the condition for the Stokes line in
the $\omega x$ complex plane to a condition in the complex $x$ plane}
\footnote{It may be worth noting that the monodromy technique requires
input about the behavior of the asymptotic frequencies from the
numerical analyses. Without the numerical results, it is not possible
to analyze the Stokes line in the $\omega x$ plane.}.

For the asymptotically flat spacetime, the numerical results
for the asymptotic QNM frequencies indicate that $\Im(\omega) \gg
\Re(\omega)$. Thus the condition for the Stokes line translates to
$\Re(x) \simeq 0$ which is nothing but the Anti-Stokes line in the
complex $x$ manifold.
\item {\it Identify a closed contour in the complex $x$ plane 
and calculate the monodromy.} 

Calculating the monodromy gives the analytical expression for the the
asymptotic QNM frequencies.
\end{enumerate}

\section{Stokes line, contours and monodromy boundary conditions}
\label{sec:mon-bc-cont}

In the previous section, we discussed main features of the monodromy
technique. In this section, we obtain the Regge-Wheeler potential in
the three regimes -- singularity, horizon(s) and spatial infinity --
and discuss generic properties of the Stokes line, contours and
monodromy boundary conditions for the three cases. [We refer to the
conditions in the $\omega x$ plane as ``monodromy'' boundary
conditions.]

\subsection{Regge-Wheeler potential}

Near the singularity, the asymptotic properties of the spacetimes
will not play any role. Hence, for the all the three cases, the
singularity structure will be identical. As mentioned earlier, we
assume that near the singularity the line-element is given by
Szekeres-Iyer metric. The generalized Regge-Wheeler potential near the
singularity is 
\ba
\label{eq:Vsing}
V \left[ r(x) \right] &\stackrel{r \to 0}{\sim}& 
\frac{q D}{8} \left( \frac{q D}{2} - 2 \right) r^{-\frac{4}{q}}
= \frac{ \left( \frac{q D}{2}-1 \right)^2-1}{4 x^2} \nn \\
&\equiv& \frac{j^2-1}{4 x^2} \, ,
\ea
\ni where
\be
\label{eq:jsing}
j = \frac{q D}{2} - 1 \, .
\ee
Substituting the potential in Eq. (\ref{eq:rweqn}), we get
(cf. Ref. \cite{Abramowitz-Steg:1964-bk}, p. 362)
\be
\label{eq:Rsing}
R(x) \sim A_+ \sqrt{2 \pi \omega x} J_{\frac{j}{2}}(\omega x) + 
A_- \sqrt{2 \pi \omega x} J_{-\frac{j}{2}}(\omega x) \, ,
\ee
where the quantities $J_{\mu}$ are the Bessel functions of order
$\mu$.  Using the asymptotic behavior of the Bessel functions
(cf. Ref. \cite{Abramowitz-Steg:1964-bk}, p. 364)
\be
\label{eq:Jasymp}
\lim_{|z| \to \infty} 
J_{\nu}(z) = \sqrt{\frac{2}{\pi z}} \cos \left( z - \frac{1}{2}\nu\pi
- \frac{1}{4}\pi \right) \, ,
\ee
we get
\be
\label{eq:Jasymp1}
\sqrt{2 \pi \omega x} J_{\pm \frac{j}{2}}(\omega x) \sim 
2 \cos \left( \omega x - \alpha_{\pm} \right) \, ,
\ee
where
\be
\label{eq:defalpha}
\alpha_{\pm} = \frac{\pi}{4}(1 \pm j) \, .
\ee
Thus, the asymptotic form of $R$ is
\ba
\label{eq:Rasymp}
R(x) &\sim& 
\left( A_+ e^{i \alpha_+} + A_- e^{i \alpha_-} \right) e^{-i \omega x} \nn \\
&+& 
\left( A_+ e^{-i \alpha_+} + A_- e^{-i \alpha_-} \right) e^{i \omega x} \, .
\ea

Near the horizons, the Regge-Wheeler potential decays exponentially
[cf. Eqs. (\ref{eq:RW-afshor},\ref{eq:RW-dshor})]. Hence, the mode
function $R(x)$ is a superposition of plane waves.

Near the spatial infinity, the potential decays (grows) for the
asymptotically flat (anti-de Sitter) spacetimes. Hence, the mode
function $R(x)$ is a superposition of plane waves (exponentially
decaying/growing solutions).  

\subsection{Stokes line}
\label{sec:stokes}

In the monodromy technique, unlike the canonical techniques, we need
to obtain solution to the differential equation (\ref{eq:rxweqn}) in
the complex plane. Thus, the canonical boundary conditions which were
defined on the boundary points in the $x$-line have to be redefined in
the $\omega x$ complex. In other words, the canonical boundary
conditions (in $x$ line) need to be mapped to the boundary conditions
on the $\omega x$ curve.

There is no unique choice for the $\omega x$ curve. In this work, we
will assume $\omega x$ to be along the {\em Stokes line}
i. e. $\Im(\omega x)=0$. There are couple of reasons for this choice:
Firstly, all along this curve the QNM solutions ($e^{\pm i \omega x}$)
are purely oscillating and do not contain any exponentially
growing/decaying modes. Secondly, using the numerical results of the
asymptotic QNM frequencies, the Stokes-line condition in the $\omega
x$ plane can be transformed onto a condition in the $x$ plane. In the
case of asymptotically flat and de Sitter cases, the Stokes line
condition translates to $\Re(x) \simeq 0$ \cite{Motl-Neit:2003} while
for asymptotically anti-de Sitter spacetimes, the Stokes line
condition translates to $\Im(\exp[i \pi/3] x) = 0$
\cite{Notario-Priv:2005}.

In the rest of this subsection, we discuss the generic properties 
of the Stokes line near the singularity and asymptotic infinity. \\

\underline{\it Near the generic singularity}: \\

Using the relation (\ref{eq:xsing}), between $x$ and $r$, near the
generic singularity, and setting
\be
\label{eq:rsing}
r = \rho e^{i \theta} \qquad {\rm where} \qquad 
\rho , \theta \in {\mathcal R} \, ,
\ee
the Stokes line condition take the following simple form: 
{\small 
\beq
\label{eq:thetasing}
\tan \left( \frac{2 \theta}{q} \right) = - \arg \omega  
\quad \Longrightarrow \quad 
\theta = - \frac{q}{2} 
\tan^{-1} \left[ \arg\omega \right] + \frac{n \pi q}{2} \, ,
\eeq
}

\ni where $n$ is an integer. This implies that (i) 
Near the generic singularity, the Stokes line have $2D$ branches and (ii)
The angle between adjacent branches are $(\pi q/{2})$.

In order to illustrate these features, we have plotted the Stokes
lines for the three cases in Figs. (1,2,3). \\

\underline{\it Near the spatial infinity/cosmological horizon}: 
\\

For the asymptotic flat spacetimes, using the fact that $x \sim r$
(near the spatial infinity), it is easy to see that the Stokes line
diverges to infinity. 

For the asymptotic de Sitter spacetimes, using the fact that $x \sim
1/r$ it is easy to see that the Stokes line cross the real axis. Thus,
the Stokes line forms a closed contour.

For the asymptotic anti-de Sitter spacetime, using the fact that $x$
is not purely real (in the asymptotic limit) it is easy to see that
Stokes line do not cross the real axis. In fact, the angle the Stokes
line makes w.r.t the real axis is $\pi/3$\footnote{Using the fact that
Stokes line condition can be rewritten as $\Im(\exp[i
\pi/3] x) = 0$ and setting $x = x_0 \exp( \pm i \theta)$, we get 
$\theta = \mp \pi/3$.}. It is also easy to see that the Stokes line do
not close in this case.

Note that the above arguments are generic and depend only on the
asymptotic properties of the spacetime. Further, as noted at the end
of Sec. (\ref{sec:nr}), the difference in the Stokes lines for the
asymptotic de Sitter/flat and anti-de Sitter confirms that the
structure of the asymptotic QNM frequencies crucially depend on the 
choice of the boundary conditions.

\subsection{Contours and monodromy}
\label{sec:contours}

In this subsection, we discuss the choice of contours (to calculate
the monodromy) for all the three cases.  All along the Stokes lines,
the modes are purely oscillating plane-waves without any exponentially
growing or decaying solutions. This property of the Stokes-lines is
useful to obtain the monodromy around a closed contour for the mode
function. Hence, we choose our contours to lie as close as possible to
the Stokes line. \\

\underline{\it Asymptotic flat spacetimes:} \\
\begin{figure}[!htb]
\begin{center}
\epsfxsize 3.00 in
\epsfysize 2.50 in
\epsfbox{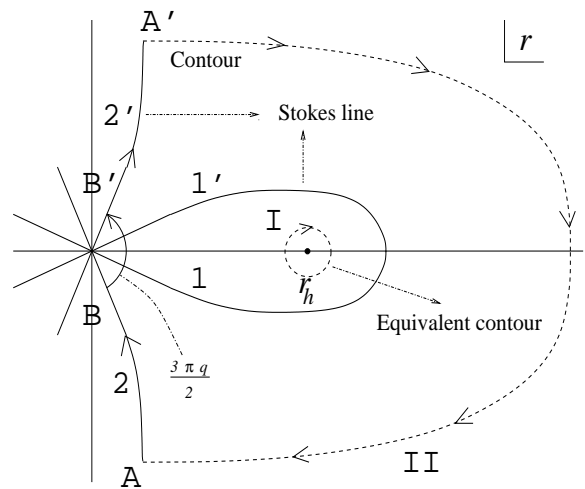}
\caption{Stokes lines and contour for asymptotically flat spacetimes.}
\label{fig:asf}
\end{center}
\end{figure}

\ni Fig. (\ref{fig:asf}) contains the contour plot for a general 
$(D + 2)-$dimensional single-horizon, asymptotically flat spherically
symmetric spacetimes. Near the singularity, the Stokes lines have $2
D$ branches. Out of these, two branches ($1$ and $1'$) of the Stokes
line go around the event horizon ($r_h$) and form a closed
contour. Two other branches ($2$ and $2'$) which extend up to infinity
do not form a closed contour. [Points $A, A'$ correspond to the points
at the spatial infinity.] Note that the angle between $2$ and $2'$ is
$(3 \pi q/{2})$.

At the spatial infinity, since the spacetime is flat, the WKB
solutions to the differential equation (\ref{eq:rweqn}) are exact
implying that the the mode function is a superposition of
plane-waves. In other words, all along the dotted line connecting the
points $A$ and $A'$ in Fig. (\ref{fig:asf}), the mode functions are
superposition of plane-waves. 

We compute the monodromy using two equivalent contours ($I, II$). The
two contours give different contributions to the mode function $R(x)$.
While contour $I$ picks up the monodromy contribution from the
horizons, contour $II$ picks up a factor from the generic singularity.
Monodromy contribution from contour $I$ is easy to evaluate while that
of contour $II$ is non-trivial and has two terms:
\br
&& \!\!\!\!\!\!\!\!\!\!\!\!\!\!\!\!\!\!
\mbox{Monodromy of the mode function}  \nn \\
= && \mbox{Factor by which coeff of $e^{\mp i \omega x}$ gets multiplied} \nn \\ 
&& \times \mbox{ Monodromy of $e^{\mp i \omega x}$} \, .
\label{eq:Tot-Mon}
\er
We then equate the monodromies obtained from contours $I$ and $II$.
The steps involved in the calculation are discussed in detail in
Sec. (\ref{sec:soln}). 

In either case, due to the logarithmic relation between $r$ and $x$
[cf. Eq. (\ref{eq:xhor})], the $e^{\mp i \omega x}$ parts of the mode
function pick up a monodromy. If in the $r$-plane we perform a
clockwise rotation around the horizon by $2 \pi$, due to discontinuity
across the branch-cut, we get
\ba
\label{eq:monfuncrh}
&& \ln(r-r_h) \to \ln(r-r_h) - 2 \pi i 
\quad \Longrightarrow \quad x \to x - \frac{\pi i}{\kappa_h} \nn \\
&& e^{\mp i \omega x} \to e^{\mp i \omega 
\left( x - \frac{\pi i}{\kappa_h} \right)} = 
e^{\mp i \omega x} e^{\mp \frac{\pi \omega}{\kappa_h}} \, .
\ea
Thus, the clock-wise rotation of the plane-wave modes in the
equivalent contour will acquire the monodromy of the following form:
\be
{\rm Monodromy} \left[ \exp({\mp i \omega x}) \right] = 
\exp\l(\mp \frac{\pi \omega}{\kappa_h}\r) \, .
\label{eq:con1-afs}
\ee 
%

\underline{\it Asymptotic de Sitter spacetimes:} \\
\begin{figure}[!htb]
\begin{center}
\epsfxsize 3.00 in
\epsfysize 2.50 in
\epsfbox{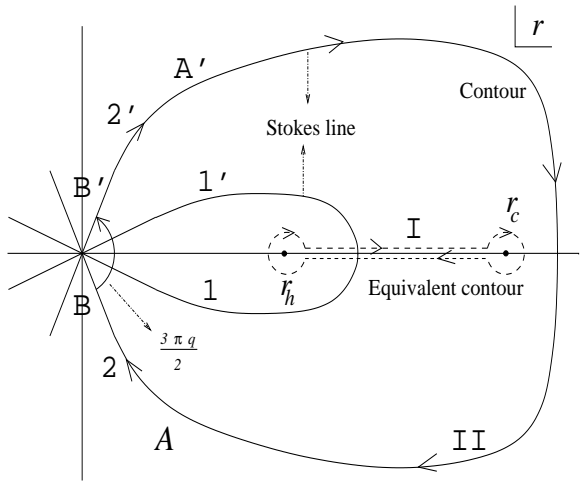}
\caption{Stokes lines and contour for asymptotically de Sitter spacetimes.}
\label{fig:ds}
\end{center}
\end{figure}

\ni Fig. (\ref{fig:ds}) contains the contour plot for a general 
$(D + 2)-$dimensional single-horizon, asymptotic de Sitter,
spherically symmetric spacetimes. As in the asymptotically flat
spacetimes, the Stokes lines have $2 D$ branches near the
singularity. Out of these, two branches ($1$ and $1'$) of the Stokes
line go around the event horizon ($r_h$) and form a closed
contour. Two other branches ($2$ and $2'$) which extend upto infinity
also form a closed contour. Note that, as in the case of
asymptotically flat spacetimes, the angle between $2$ and $2'$ is $(3
\pi q/{2})$. In order to evaluate the monodromy around the contour
this is the angle by which we need to deform the contour close to the
singularity.

The procedure to compute the monodromy is similar to that of the
asymptotically flat spacetime case except that in this case the
monodromy has to evaluated for two -- event and cosmological --
horizons. The total monodromy of the mode functions is again given by
the relation (\ref{eq:Tot-Mon}). 

Here again, the monodromy contribution from contour $I$ is easy to
evaluate while that of contour $II$ is non-trivial
[Eqn. (\ref{eq:Tot-Mon})] and we equate the two monodromies from the
two contours. The details are discussed in Sec. (\ref{sec:soln}).  In
this case, the contours go clockwise around both the horizon and the
cosmological horizon. Thus, the plane-wave modes $e^{\mp i \omega x}$
pick up the following monodromy term
\be
{\rm Monodromy} \left[ \exp({\mp i \omega x}) \right] = 
\exp\l(\mp \frac{\pi \omega}{\kappa_h} \mp \frac{\pi \omega}{\kappa_c} \r) \, .
\label{eq:con1-ds}
\ee 
%

\underline{\it Asymptotic Anti-de Sitter spacetimes:} \\
\begin{figure}[!htb]
\begin{center}
\epsfxsize 3.00 in
\epsfysize 2.50 in
\epsfbox{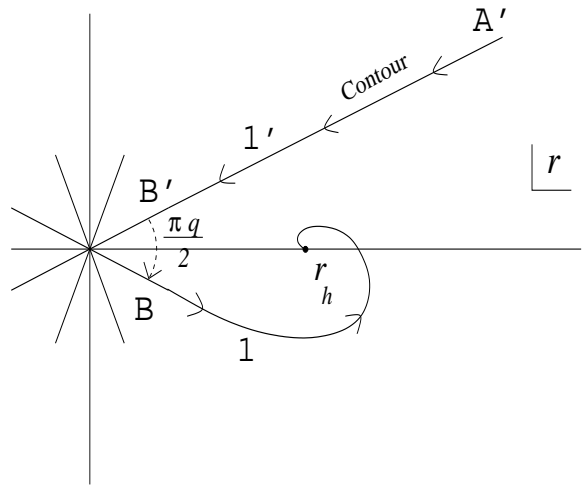}
\caption{Stokes lines and contour for asymptotically Anti-de Sitter spacetimes.}
\label{fig:ads}
\end{center}
\end{figure}

\ni Fig. (\ref{fig:ads}) contains the contour plot for a general 
$(D + 2)-$dimensional single-horizon, asymptotic anti-de Sitter,
spherically symmetric spacetimes. As in the previous two cases, the
Stokes lines have $2 D$ branches near the singularity. Utilizing the
multi-valuedness of the tortoise coordinate near the horizon, we choose
the Stokes line from the horizon to join one branch ($1'$) of the
Stokes line from the origin. Another branch ($1$), from the origin, is
chosen to extend upto infinity. 

Unlike the previous two cases, even one of the Stokes line (hence
contour) do not close. Note that the angle between $1$ and $1'$ is
$(\pi q/2)$. This is the angle by which we need to deform the contour
close to the singularity.

Since the contour does not close, we can not use the monodromy
technique.  In this case, we obtain the high-frequency QNM by matching
the asymptotic solutions with the exact solutions in the limit of
$\omega x -to \pm \infty$.

\subsection{Monodromy boundary conditions}
\label{sec:mono}

For highly damped modes, 
\beq
x \to \pm \infty \qquad \Longrightarrow \qquad 
\omega x \to \pm i \infty \, .
\eeq
Thus, the boundary points in the real $x$-line are purely imaginary in
the $\omega x$ plane [points $A, B$ in
Figs.(\ref{fig:asf},\ref{fig:ds})]. Using the fact that the
``canonical'' boundary conditions are identical for the asymptotically
flat and de Sitter spacetimes and that the contours are similar for
the two cases, the ``monodromy'' boundary conditions are given
by\footnote{There are some subtleties involved in defining the
boundary condition at the spatial infinity. See
Ref. \cite{Motl-Neit:2003}.}
\be
\label{eq:Mono-BC}
R(x) \sim e^{\mp i \omega x} 
\qquad \left\{ 
\begin{array}{lll}
\omega x \to \pm \infty & & \Re(\omega)>0 \\
\omega x \to \mp \infty & & \Re(\omega)<0
\end{array}
\right. \, .
\ee

In the case of asymptotic anti-de Sitter spacetimes, it is not
possible to map the ``canonical'' boundary conditions on to the
$\omega x$ plane.

\section{Computing the asymptotic QN frequencies}
\label{sec:soln}

In this section, we compute the asymptotic QN frequencies for the
general $(D + 2)$-dimensional spherically symmetric spacetimes .  In
the following subsection, using the monodromy technique, we compute
the high frequency QNMs for the asymptotically flat and de Sitter
spacetimes in a unified manner . In the last subsection, we obtain
the high QNM frequencies for the asymptotic anti-de Sitter
spacetimes.

\subsection{Asymptotic flat and de Sitter spacetimes}
\label{sec:flat-dS}

We first compute the monodromy contribution from contour $II$. In
order to do that, we follow the contour from the negative imaginary
axis ($A$) to the positive imaginary axis ($A'$) -- by passing through
the points $B$ and $B'$ -- and come back to $A$.

The mode function $R(x)$ at $A$ is given by (\ref{eq:Rasymp}). In the
case of asymptotically flat spacetimes, as in the previous analyses,
it is possible to fix the constants -- by applying the boundary
conditions (\ref{eq:Mono-BC}) -- before calculating the
monodromy. However, we would like to follow a different procedure: we
apply the boundary conditions after calculating the monodromy. In this
way, it is possible to obtain the high-frequency QNMs for the
asymptotic flat and de Sitter spacetimes in a unified manner.

At $B$, the mode function $R(x)$ is given by (\ref{eq:Rsing}).  To
obtain the mode function $R(x)$ at $B'$, we need to deform the contour
close to the singularity (in the $r$-plane) by an angle $(3 \pi
q/{2})$ . Using Eq. (\ref{eq:xsing}), this translates to a rotation by
an angle $(3 \pi q/{2}) \times (2/{q}) = 3 \pi$ in the $x$ plane i.e.
$x \to x^{\prime} = e^{i 3 \pi}x$.

Using the relation
\be
\label{eq:Jrotn}
J_\n(z e^{im\p}) = e^{im\n\p} \, J_\n(z) \, ,
\ee
we get
\beq
\label{eq:rotn3pi}
\sqrt{\omega x^{\prime}} J_{\pm \frac{j}{2}}(\omega x^{\prime}) 
= e^{i 6 \alpha_{\pm}} \sqrt{\omega x} J_{\pm \frac{j}{2}}(\omega x) \, . 
\eeq
Thus, the mode function $R(x)$ at $B'$ is given by 
\br
R(x) &\sim& A_{+} \sqrt{2 \pi \omega x} \, \exp(i 6 \alpha_{+}) 
J_{\frac{j}{2}}(\omega x) \nn \\
&+& A_- \sqrt{2 \pi \omega x} \, \exp(i 6 \alpha_{-})  
J_{-\frac{j}{2}}(\omega x) \, .
\er
Using the asymptotic expansion (\ref{eq:Jasymp}), the mode function at
$A'$ is given by
\ba
\label{eq:Rasymp3pi}
R(x) &\sim& \left( A_+ e^{5 i \alpha_+} + 
A_- e^{5 i \alpha_-} \right) e^{-i \omega x} \nn \\
&+& \left( A_+ e^{7 i \alpha_+} + 
A_- e^{7 i \alpha_-} \right) e^{i \omega x} \, .
\ea

As we go along the closed contour from $B$ to $A$, the ``new'' mode
function (\ref{eq:Rasymp3pi}) is different compared to the original
mode function (\ref{eq:Rasymp}). The coefficients of $e^{\mp i \omega
x}$ in the mode function $R(x)$ are different by a factor
\ba
\label{eq:factor1}
\frac{A_+e^{5 i \alpha_+} + A_-e^{5 i \alpha_-}}{A_+e^{i \alpha_+} 
+ A_-e^{i \alpha_-}} \, \,\, {\rm and} \, \, \, 
\frac{A_+e^{7 i \alpha_+} + A_-e^{7 i \alpha_-}}{A_+e^{-i \alpha_+} 
+ A_-e^{-i \alpha_-}} \, ,
\ea
\ni respectively. 

The monodromies of the components $e^{\mp i \omega x}$ are given by
Eqs. (\ref{eq:con1-afs},\ref{eq:con1-ds}).  In order to compute the
monodromy contribution from contour $I$, we notice that the contour
passes close to the horizon and the cosmological horizon. There, the
mode functions are purely ingoing and outgoing functions, $e^{i \omega
x}$ and $e^{- i \omega x}$ respectively. Therefore, the monodromy of
the mode function is the monodromy of $e^{i \omega x}$ at the horizon
and that of $e^{- i \omega x}$ at the cosmological horizon. The total
monodromy is
\br
\label{eq:tot-ds}
\mbox{\small asymptotic de Sitter} &:& 
\exp\l[{\pi \omega(\frac{1}{\kappa_h} - \frac{1}{\kappa_c})}\r] \\ 
\label{eq:tot-afs}
\mbox{\small asymptotic flat} &:& 
\exp\l[\frac{\pi \omega}{\kappa_h}\r] \, .
\er
Substituting expressions
(\ref{eq:con1-afs},\ref{eq:con1-ds},\ref{eq:factor1},\ref{eq:tot-ds},\ref{eq:tot-afs})
in Eq. (\ref{eq:Tot-Mon}), we get, for asymptotic de Sitter spacetimes
\begin{subequations}
\label{eq:constraint-ds}
\ba
\label{eq:constraint-dsa}
\frac{A_+e^{5 i \alpha_+} + A_-e^{5 i \alpha_-}}{A_+e^{i \alpha_+} 
+ A_-e^{i \alpha_-}} &=& e^{\frac{2 \pi \omega}{\kappa_h}} \, , \\
\label{eq:constraint-dsb}
\frac{A_+e^{7 i \alpha_+} + A_-e^{7 i \alpha_-}}{A_+e^{-i \alpha_+} 
+ A_-e^{-i \alpha_-}} &=& e^{-\frac{2 \pi \omega}{\kappa_c}} \, .
\ea
\end{subequations}
[For completeness, we have given the detailed derivation of the above
result in Appendix (\ref{app:two}).]

In order to obtain the asymptotically flat results, we set $\kappa_c
\to 0^-$. Eqn. (\ref{eq:constraint-dsa}) is independent of $\kappa_c$
and remains unaffected. The RHS of (\ref{eq:constraint-dsb}) either grows 
or decays exponentially depending on whether $\Re(\omega)$ is positive or
negative.  Thus, for the asymptotically flat case, we obtain
\begin{subequations}
\label{eq:constraint-af}
\ba
\label{eq:constraint-afa}
\!\!\!\!\!\!\!\!\!
\frac{A_+e^{5 i \alpha_+} + A_-e^{5 i \alpha_-}}{A_+e^{i \alpha_+} 
+ A_-e^{i \alpha_-}} &=& e^{\frac{2 \pi \omega}{\kappa_h}} \, , \\
\label{eq:constraint-afb}
A_+ e^{-i \alpha_+} + A_- e^{-i \alpha_-} &=& 0 \ \ \ \ (\Re(\omega)>0), \\
\label{eq:constraint-afc}
A_+ e^{7 i \alpha_+} + A_- e^{7 i \alpha_-} &=& 0 \ \ \ \ (\Re(\omega)<0).
\ea
\end{subequations}
Note that the constraint (\ref{eq:constraint-afb}) is same as that
obtained in Ref. \cite{Motl-Neit:2003}. However, the constraint we
obtain in Eq. (\ref{eq:constraint-afc}) is new. As we will see
below, this constraint gives the correct asymptotic QNM frequency with
$\Re(\omega_{QNM}) < 0$. In all the earlier analysis, in order to
obtain the second set of asymptotic QNM frequencies ($\Re(\omega_{QNM})
< 0$), the authors run the contour in the opposite direction. In our
formalism, this emerges naturally. 

Eliminating $A_+$, $A_-$ from
(\ref{eq:constraint-ds},\ref{eq:constraint-af}),
we obtain the
expression for the asymptotic QNM frequencies for the two cases:
{\small
\ba
\label{eq:finres-ds}
& & \tanh \left( \frac{\pi \omega_{_{QNM}}}{\kappa_h} \right) 
\tanh \left( \frac{\pi \omega_{_{QNM}}}{\kappa_c} \right)
= \frac{2}{\tan^2 \left[ \frac{\pi\left(q D -2\right)}{4} \right] - 1 } 
\, , \, \, \, \,\, \, \, \, \, \, \, \, \\
& & \qquad \qquad \qquad \mbox{(asymptotic de Sitter spacetimes)} \nn \\
\label{eq:finres-afs}
& & \frac{2 \pi \omega_{_{QNM}}}{\kappa_h} = 
\left( 2n+1 \right) i \pi \pm \log 
\left[ 1 + 2 \cos \l(\frac{\pi \left(q D -2\right)}{2}\r) \right ] \, . \\
& & \qquad \qquad \qquad \mbox{(asymptotically flat spacetimes)} \nn
\ea
}

\ni We would like to stress the following points regarding the above
results: First, the above results are valid for a single
event-horizon, spherically symmetric spacetimes which are
asymptotically flat and de Sitter. Second, we have obtained the
asymptotic QNM frequencies for the two cases in a unified
manner. Third, the expression for the asymptotic de Sitter
spacetimes -- unlike the flat spacetime -- is transcendental,
hence it is not possible to obtain solutions uniquely. 
Fourth, in the case of asymptotic de Sitter spacetimes our
result matches with that of Natario and Schiappa
\cite{Natario-Schi:2004} for the specific case $qD = 2$. Fifth, 
in the limit of $\kappa_c \rightarrow 0^-$, Eq. (\ref{eq:finres-ds})
gives the expression for the asymptotically flat spacetimes
(\ref{eq:finres-afs}). In the limit of $k_h \to 0^+$,
Eq. (\ref{eq:finres-ds}) gives
\beq
\frac{\omega_{_{QNM}}}{\kappa_c} =  - \frac{1}{2\pi} \ln(3) \pm 
i \l(\frac{1}{2} + n\r) \, ,
\eeq
which is not the same for the exact de Sitter. As shown earlier
\cite{Natario-Schi:2004}, the asymptotic de Sitter limit does not
provide the correct limit for pure de Sitter. Lastly, in the case of
asymptotically flat spacetimes, even though the condition
(\ref{eq:constraint-afc}) is different compared to that obtained in
the earlier analyses, the expression for the high QNM frequencies are
exactly the same.

\subsection{Asymptotic Anti-de Sitter spacetimes}
\label{sec:AdS}

In this case we can not calculate the monodromy, since the Stokes line do
not form a closed contour. Instead, we do the following: We calculate
the exact mode function near the generic singularity ($B'$), horizon
($r_h$) and spatial infinity ($A'$). We find the asymptotic limit of
the modes along the branches $1$ and $1'$. [Note that the asymptotic
limit corresponding to $\omega x \to \infty$ corresponds to spatial
infinity while $\omega x \to - \infty$ corresponds to the event
horizon.]  Matching the asymptotic solutions with the exact solutions
at $A'$ and $r_h$, we obtain the analytic expression for the high
QNM frequencies. [We follow the notation of
Ref. \cite{Natario-Schi:2004} closely to provide easy comparison.]

First, we match the asymptotic and exact mode function at spatial
infinity by going along branch $1$. The exact mode functions at point
$B'$ is given by Eq. (\ref{eq:Rsing}). The asymptotic limit
corresponding to point $A'$ i. e. $\omega x \to \infty$ is given by
Eq. (\ref{eq:Rasymp}).

The exact mode function at spatial infinity $A'$ is given by
Eq. (\ref{eq:Rinfty-bcads}). Using the relations
[cf. Ref. \cite{Abramowitz-Steg:1964-bk}]
\br
J_{n}(z) &\sim& \l(\frac{2}{\pi z}\r)^{1/2} \cos\l(z - \frac{n \pi}{2} - 
\frac{\pi}{4}\r) \\
J_{n + 1/2}(z) &=& \l(\frac{2}{\pi}\r)^{1/2} z^{n + 1/2} 
\l(- \frac{d}{z dz}\r)^n \l(\frac{\sin z}{z}\r) \nn \, ,
\er
(for odd and even dimensions respectively),
Eq. (\ref{eq:Rinfty-bcads}) can be rewritten as
\beq
\label{eq:Rasympads1}
R(x) =  B_+ \left[ e^{-i \beta_+} e^{i \omega x_0} e^{-i \omega x} 
+ e^{i \beta_+} e^{-i \omega x_0} e^{i \omega x} \right] \, ,
\eeq
where 
\beq
\beta_{+} = \frac{\pi}{4} \l(1 + j_{\infty}\r) \, .
\eeq
Comparing the coefficients of $e^{\pm i \omega x}$ in the expressions
(\ref{eq:Rasymp}, \ref{eq:Rasympads1}), we  get the first constraint 
equation: 
\be
\label{eq:constraint1ads}
\frac{A_+ e^{i \alpha_+} + A_- e^{i \alpha_-}}
{A_+ e^{-i \alpha_+} + A_- e^{-i \alpha_-}} = 
\frac{e^{-i \beta_+} e^{i \omega x_0}}{e^{i \beta_+} e^{-i \omega x_0}} \, .
\ee

Having obtained the first constraint, our next step is to match the
exact and asymptotic modes at the horizon by going along the branch
$1'$. In order to do that, we need to know the exact mode function at
$B$.  To obtain the mode function $R(x)$ at $B$, we need to deform the
contour close to the singularity (in the $r$-plane) by an angle $(-
\pi q/{2})$ . From (\ref{eq:xsing}), this translates to a rotation by
an angle $(- \pi q/{2}) \times (2/{q}) = - \pi$ in the $x$ plane i.e.
$x \to x^{\prime} = e^{i 3 \pi}x$. Using the relation (\ref{eq:Jrotn}), 
we get 
\br
R(x) &\sim& A_{+} \sqrt{2 \pi \omega x} \, \exp(-i 3 \alpha_{+}) 
J_{\frac{j}{2}}(\omega x) \nn \\
&+& A_- \sqrt{2 \pi \omega x} \, \exp(-i 3 \alpha_{-})  
J_{-\frac{j}{2}}(\omega x) \, .
\er
The asymptotic limit of the above mode functions (corresponding to
$r_h$) reduces to the following simple form:
\ba
\label{eq:Rasymp-pi}
R(x) &\sim& \left( A_+ e^{-3 i \alpha_+} 
+ A_- e^{-3 i \alpha_-} \right) e^{-i \omega x} \nn \\ 
&+& \left( A_+ e^{-i \alpha_+} +
 A_- e^{-i \alpha_-} \right) e^{i \omega x} \, .
\ea
Comparing the coefficients of $\e^{\pm i \omega x}$ in the expressions
(\ref{eq:bc2a}, \ref{eq:Rasymp-pi}), we get the second constraint
equation: 
\be
\label{eq:constraint2ads}
A_+ e^{-3 i \alpha_+} + A_- e^{-3 i \alpha_-} = 0 \, .
\ee
Eliminating $A_+$, $A_-$ from
(\ref{eq:constraint1ads},\ref{eq:constraint2ads}), we obtain the
analytical expression for the asymptotic QNM frequencies for the asymptotic 
anti-de Sitter spacetimes:
\beq
\label{eq:resultads}
\omega_{_{QNM}} x_0 = \frac{\pi}{2} \l(2n + \frac{D + 3}{2} \r) 
- \frac{i}{2} \log \left[ 2 \cos \frac{\pi\left(qD-2\right)}{4}
\right] \, .
\eeq
[For continuity, we have given the detailed derivation of the above
result in Appendix (\ref{app:two}).] 

We would like to stress the following points regarding the above
result: First, the above result is valid for a general, single-horizon
spherically symmetric asymptotic anti-de Sitter spacetimes. Second,
unlike the asymptotic flat spacetime, the high QNM frequencies has no
generic features. This is because $x_0$ is a arbitrary complex number
which depends on the properties of the spacetime. Third,
$\omega_{_{QNM}} x_0$ is purely real when $q D = 10/3$. In the case of
$D = 2, q = 1$, we get
\be
\label{eq:resultadsD=2}
\omega_{_{QNM}} x_0 = \left( n + \frac{1}{4} \right) \pi 
- \frac{i}{2} \log 2
\ee
Lastly and more importantly, it is clear from the
above expression that the real and imaginary parts of the QNM
frequency are of similar order unlike the asymptotic flat/de Sitter
cases.

\section{Application to specific black holes}
\label{sec:specificBH}

In the previous section, we obtained master equations for the high
frequency QNM for a spherically symmetric black hole with a generic
singularity with three different asymptotic properties. As we have
shown, the real part of the high frequency QNM is not necessarily
proportional to $\ln(3)$ as in the case of $(D + 2)$-dimensional
Schwarzschild. In order to illustrate this fact, we take specific
examples and obtain their QNM.

\subsection{$(D+2)-$dimensional Schwarzschild-de Sitter}

In the case of $(D + 2)-$dimensional Schwarzschild-de Sitter, the
functions $f(r), g(r)$ and $\rho(r)$ in the line-element
(\ref{eq:spher-tr}) are given by
\beq
f(r) = g(r) = 1 - \left(\frac{r_h}{r}\right)^{D - 1} + \frac{r^2}{\ell^2} ~; 
~ \rho(r) = r \, ,
\label{eq:line-sds}
\eeq
where $r_h$ is related to the black hole mass $(M)$ and the $(D +
2)-$dimensional cosmological constant $\Lambda$.

Comparing Eqs. (\ref{eq:fg-r0},\ref{eq:line-sds}), we get
\be
p = \frac{1 - D}{D}~;\quad  q = \frac{2}{D}~; \quad x = \f{r^D}{\b}~,
\label{eq:sds-pqx}
\ee
Substituting the above expressions Eq. (\ref{eq:finres-ds}), we get
\beq
\tanh \left(\pi \omega_{_{QNM}}/\kappa_h\r)
\tanh \left(\pi \omega_{_{QNM}}/\kappa_c\r) = - 2 \, .
\label{eq:ome-Sds}
\eeq
The above expression matches with that obtained by the previous authors 
\cite{Natario-Schi:2004}. 

\subsection{$(D+2)-$dimensional Schwarzschild Anti-de Sitter}

In the case of $(D + 2)-$dimensional Schwarzschild anti-de Sitter, the
functions $f(r), g(r)$ and $\rho(r)$ in the line-element
(\ref{eq:spher-tr}) are given by
\beq
f(r) = g(r) = 1 - \left(\frac{r_h}{r}\right)^{D - 1} - \frac{r^2}{\ell^2} ~; 
~ \rho(r) = r \, ,
\label{eq:line-sads}
\eeq
where $r_h$ is related to the black hole mass $(M)$ and the $(D +
2)-$dimensional cosmological constant.

Near the singularity, the structure of the metric is same as that 
of the Schwarzschild-de Sitter. Thus, near the singularity the 
expressions remain the same [cf. (\ref{eq:sds-pqx})].

Substituting the above expressions Eq. (\ref{eq:resultads}), we get
\beq
\omega_{_{QNM}} x_0 = \frac{\pi}{2} \l(2n + \frac{D + 3}{2} \r) 
- \frac{i}{2} \log(2) \, .
\label{eq:ome-Sads}
\eeq
Even though the above expression is valid of Schwarzschild-anti de
Sitter black holes it is, in general, not possible to obtain a
explicit expression for the $x_0$ since it is a complicated function
of $M$ and $\ell^2$. It is possible to obtain a closed expression of
$x_0$ only in the large black hole limit $({\rm horizon~radius}/\ell \ll
1)$:
\beq
x_0 = \frac{\pi}{2 \kappa_h} \frac{\exp[-i \pi/(D + 1)]}{\sin[-i \pi/(D + 1)]} 
\eeq

\subsection{Non-rotating BTZ black hole} 

The line-element of the $3$-dimensional non-rotating BTZ black hole
\cite{Banados-Henn:1992a} is 
\beq
 ds^2 = -\l( \frac{{r}^2}{\ell^2}  - M \r ) d t^2 + 
\l(\frac{{r}^2}{\ell^2} - M \r)^{-1} \!\!\! d{r}^2 + {r}^2 d{\varphi}^2 \, , 
\label{eq:line-btz}
\eeq
where $M$ is the mass of the black hole and $\ell^2$ is related to the
negative 3-d cosmological constant. The above solution has an event
horizon at $\ell \sqrt{M}$ while there is no singularity at the
origin.

Even though the BTZ line-element does not have a singularity at the
origin, it is possible to compare the line-element with that of the
generic singularity (\ref{eq:gensing1}). We get
\be
p = 0~;\quad  q = 2; \quad x \sim r \, .
\label{eq:btz-pqx}
\ee
Substituting the above expressions in Eq. (\ref{eq:resultads}), we get
\beq
\omega_{_{QNM}} = - 2 i \sqrt{M} \l(n + 1\r) + \frac{\log 2}{\pi}  \, .
\eeq
The real part in the RHS of the above expression does not match with
that of the earlier analyses (cf. Ref. \cite{Cardoso-Phd:2004}). In 
all the earlier analyses, the real part is equal to $l$ (constant).

The reason for the discrepancy is as follows: (i) Our analysis, rests
on the fact that the spacetime has a singularity at the
origin. However, the BTZ black hole does not have a singularity. (ii)
For BTZ $p - q + 2 = 0$, hence the second condition in
Eq. (\ref{eq:pq-cond}) is violated. This implies that the near the
origin, the dominant term in the Regge-Wheeler potential is not given
by Eq. (\ref{eq:Vsing}) and by the following expression:
\br
V[r(x)] &\stackrel{r \to 0}{\sim}&  \frac{l^2}{r^2} * M 
\er
These suggest that the naive application of our formalism does not 
work.

\section{Discussion and Conclusion}
\label{sec:disc}

In this work, we have computed the high frequency QNMs for scalar
perturbations of spherically symmetric single horizons in
$(D+2)-$dimensional -- asymptotically flat, de Sitter and anti-de
Sitter -- spacetimes. We have computed these modes using the
monodromy approach \cite{Motl-Neit:2003}. In all the three cases, we
have shown that the asymptotic frequency of these modes depends on the
surface gravity of the event horizon ($\kappa_h$), the cosmological
constant ($\Lambda$), dimension ($D$) and the power-law index ($q$) of
$S^D$ near the singularity.

Unlike the earlier analyses, we have computed the high-frequency QNMs
for the asymptotically flat and de Sitter spacetimes in a unified
manner. We have shown that: (i) In the case of asymptotic flat
spacetimes, the real part of the high frequency modes has a
logarithmic dependence, although the argument of the logarithm is not
necessarily an integer.  (ii) In the case of asymptotic non-flat
spacetimes, the real part of the high-frequency modes, in general, do
not have a lograthmic dependence. We have also applied our results to
specific examples. In the case of $(D + 2)-$dimensional Schwarzschild
de Sitter and anti-de Sitter spacetimes, our results match with that
of Natario and Schiappa \cite{Natario-Schi:2004}. However, the naive
application of our formalism does not work for the non-rotating BTZ
black hole. This is due to the fact that the BTZ black hole is
non-singular at the center.

The analysis differs from that of the earlier analyzes in two ways:
Firstly, using our analyis, a universal feature seems to emerge on the
dependence of the high QNM frequencies. It is clear from Eqs.
(\ref{eq:finres-ds}, \ref{eq:finres-afs}, \ref{eq:resultads}) that the
asymptotic QNM frequencies depend on the power-law index $q$ and {\it
  not} $p$. More importantly, the high QNM frequencies seem to have
universal dependence of the form $(D q - 2)/2$. Such a feature does
not emerge from the previous analyses especially from that of Natario
and Schiappa \cite{Natario-Schi:2004}. Secondly, our analysis can be
extended to the time-dependent black-holes. In such a case, the
generalized Regge-Wheeler potential (\ref{eq:rwpot}) will be
time-depedent. Recently, Xue etal \cite{Xue-Shen:2003} have
numerically obtained the QNM frequencies for 4D Schwarzschild. They
showed that the QNM frequencies change due to the time-depedence. It
will be interesting to analyze their results for the generic
spherically symmetric space-times. 
  
In the light of the above results, let us re-examine Hod's conjecture,
which rests on the fact that black hole entropy $S_{BH}$ is equispaced
and the number of black hole states $\Omega=\exp(S_{BH})$ is an
integer.  Consider the adiabatic invariant: \be I = \int
\frac{dE}{\omega_{QNM}}~.  \ee In the case of flat spacetimes, it
turned out generically that $\omega_{QNM} \propto T_H$ [{\bf I}]. From
this and the first law of black hole thermodynamics, one obtains:
\be
I  \propto S_{BH}~,
\ee
where $S_{BH}$ is the black hole entropy.  Since adiabatic invariants
are supposed to be equispaced, it follows that $S_{BH} \propto n$, an
integer. Further, if the proportionality constant is of the form
$\ln(\rm{interger})$, then $\Omega$ is an integer. In case of
asymptotically de Sitter spacetimes however, there is no closed form
algebraic expression for the asymptotic QNM frequencies. Thus, it is
not clear whether $S_{BH}$ is equispaced, and the degeneracy an
integer. For asymptotically anti-de Sitter spacetimes, although
$\omega_{QNM}$ can be expressed in a closed form, the latter depends
on the undetermined quantity $x_0$. Hence, once again, it seems
unlikely that $S_{BH}$ is equispaced and $\Omega$ an integer. Thus, it
appears that properties which held for asymptotically flat spacetimes,
may no longer hold under more general circumstances.


In order for the asymptotic QNM frequencies to be related to the
black hole entropy, the following quantity needs to be an adiabatic
invariant (see, for example, Ref. \cite{Kunstatter:2002}):
\beq
I = \int \f{dE}{\o_{QNM}} \, . \nn 
\eeq
The crucial ingredient in order to show that $I$ is an adiabatic
invariant is $\Re(\omega_{_{QNM}})\propto T_H$. Although it is straight
forward to show that $I$ is indeed an adiabatic invariant in the case
of asymptotic flat spacetimes, however it is far from obvious 
(and in the worst scenario, not true,) for the asymptotic non-flat
spacetimes. In the case of asymptotic de Sitter spacetimes since
there is no algebraic solution for the asymptotic QNM frequencies, it
is not possible to show that $I$ is an adiabatic invariant. In the
case of asymptotic anti-de Sitter spacetimes, althought the algebraic
structure exists the asymptotic modes crucially depend on the form of
$x_0$ (whose form is not known), hence it is again not possible to
show, in general, $I$ is an adiabatic invariant.

Thus, in the case of asymptotically non-flat spacetimes, using
high-frequency QNMs it is not possible to confirm Bekenstein's
conjecture that horizon area is an adiabatic invariant implying that
the Hod's conjecture may be restrictive.

\section*{Acknowledgments}

This work was supported by the Natural Sciences and Engineering
Research Council of Canada. We would like to thank V. Cardoso and
J. M. Natario for useful e-mail correspondences. AG and SS would like
to thank the Department of Physics, University of Lethbridge, Canada
for hospitality where most of this work was done.

\appendix

\section{Asymptotic AdS solutions}
\label{app:Sol-AdS}

In this appendix, we obtain the solution to the Regge-Wheeler equation
(\ref{eq:rweqn}) for the asymptotic AdS spacetimes. We follow closely
the approach given by Ref.  \cite{Winstanley:2001}. For these
spacetimes, we have $f(r)=g(r)$ and $\rho(r)=r$.

Substituting the Regge-Wheeler potential (\ref{eq:RW-ads1}) in
Eq. (\ref{eq:rweqn}), we get,
\ba
\label{eq:Rinftyads}
\!\!\!R(x) &\sim& B_+ \sqrt{2 \pi \omega (x_0-x)} 
J_{\frac{j_\infty}{2}} \left(\omega (x_0-x)\right) \nn \\
&+& B_- \sqrt{2 \pi \omega (x_0-x)} J_{-\frac{j_\infty}{2}}
\left(\omega (x_0-x)\right) \, ,
\ea
where $B_{\pm}$ are the constants of integration determined by the
boundary conditions and the quantities $J_{\mu}$ are the Bessel
functions of order $\mu$. Using the Bessel form for small arguments
[cf. Ref. \cite{Abramowitz-Steg:1964-bk}, p. 360]
\be
\label{eq:Jlim0}
J_\nu (x) \sim x^\nu \qquad {\rm as} \qquad x \to 0
\ee
we get
\br
& & 
\sqrt{\omega (x_0-x)} J_{-\frac{j_\infty}{2}} \left[\omega (x_0-x)\right] 
\sim (x_0-x)^{-\frac{D}{2}} \to \infty \nn \\
& &  \sqrt{\omega (x_0-x)} J_{\frac{j_\infty}{2}} 
\left[\omega (x_0-x)\right] \to 0 \, .
\er
Thus, in the spatial infinity, one of the mode function blows up while
the other decays.

In order to see things more transparently, let us perform a coordinate
transformation such that the Regge-Wheeler potential remains finite at
infinity. Introducing the following transformation:
\be
\label{eq:winr}
\tilde{r} = \log \left( r - r_h \right) \, , \,
R(r) = \frac{\left( r - r_h
\right)^\frac{1}{2}}{\sqrt{f}}\tilde{R}(\tilde{r}) \, ,
\ee
we get
\be
\label{eq:winrweqn}
\partial_{\tilde{r}}^2 \tilde{R}(\tilde{r}) - \tilde{V}[r(\tilde{r})] \tilde{R}(\tilde{r}) = 0 \, ,
\ee
where
\ba
\label{eq:winrwpot}
&& \tilde{V}(r) = -\frac{\left( r - r_h \right)^2}{f^2}\omega^2 
+ \frac{\left( r - r_h \right)^2}{f} \frac{l \left( l+D-1 \right)}{r^2} \nn \\ 
&&  - \frac{\left( r-r_h \right)^\frac{3}{2}}{\sqrt{fr^D}} 
\frac{d}{dr}\left[ f \, r^D \frac{d}{d r} 
\left\{ \frac{\left( r-r_h \right)^\frac{1}{2}}{\sqrt{fr^D}} \right\} 
\right] \nn \\
\ea
and $r$ is understood to be $r(\tilde{r})$. In the spatial infinity  
($f(r) \sim |\Lambda|r^2$), we get 
\be
\label{eq:winvasymp}
\tilde{V}(r) \stackrel{r \to \infty}{\sim} \frac{\left( D+1 \right)^2}{4} \, .
\ee
It is easy to note that Regge-Wheeler potential is positive definite
at infinity leading to a pair of exponential solutions:
\beq
\label{eq:winexp}
\tilde{R}(\tilde{r}) \sim \exp\l(\pm \frac{D+1}{2}  \tilde{r} \r) \, .
\eeq
Using (\ref{eq:winr}), we get,
\br
\label{eq:winradialasymp}
\frac{R(r)}{r^\frac{D}{2}} &\stackrel{r \to \infty}{\sim} & 
 r^{\l(-\frac{D+1}{2} \pm \frac{D + 1}{2}\r)} \nn \\
R(r) &\stackrel{r \to \infty}{\sim} & C_1^{+} r^\frac{D}{2} 
+ C_2^{+} r^{-(\frac{D}{2} + 1)} \, .
\er
The last expression is identical to Eq. (\ref{eq:plawav2}) in
Sec. (\ref{sec:bc}). $C_1^{+} = 0$ (which also implies $B_{-} = 0$)
corresponds to the reflection boundary conditions. Thus, the exact 
mode function at spatial infinity with the reflection boundary 
condition is given by 
\beq
\label{eq:Rinfty-bcads}
R(x) \sim B_+ \sqrt{2 \pi \omega (x_0-x)} 
J_{\frac{j_\infty}{2}} \left(\omega (x_0-x)\right) \\
\eeq
%


\section{Calculations}
\label{app:two}

In this appendix, we outline the essential steps leading to the master 
equations (\ref{eq:finres-afs}, \ref{eq:finres-ds}, \ref{eq:resultads}). 

\subsection{Asymptotic flat and de Sitter spacetimes}

Substituting expressions
(\ref{eq:con1-afs},\ref{eq:con1-ds},\ref{eq:factor1},\ref{eq:tot-ds},\ref{eq:tot-afs})
in Eq. (\ref{eq:Tot-Mon}), for the asymptotic de Sitter and asymptotic
flat spacetimes, we get,
{\small
\ba
\frac{A_+e^{5 i \alpha_+} + A_-e^{5 i \alpha_-}}{A_+e^{i \alpha_+} 
+ A_-e^{i \alpha_-}} \times 
e^{-\pi \omega(\frac{1}{\kappa_h}+\frac{1}{\kappa_c})} 
&=& e^{\pi \omega(\frac{1}{\kappa_h}-\frac{1}{\kappa_c})} \, \, \, \, \, \, \, \, 
\label{eq:constraint1}\\
\frac{A_+e^{7 i \alpha_+} + A_-e^{7 i \alpha_-}}{A_+e^{-i \alpha_+} 
+ A_-e^{-i \alpha_-}} \times e^{\pi \omega(\frac{1}{\kappa_h}+ 
\frac{1}{\kappa_c})} &=& 
e^{\pi \omega(\frac{1}{\kappa_h}-\frac{1}{\kappa_c})} \, \, \, \, \, \, \, \,
\label{eq:constraint2}
\ea
}
\ni
Simplifying the above expressions we get (\ref{eq:constraint-ds}). 

Eliminating $A_{\pm}$ in Eq. (\ref{eq:constraint-ds}), for the
asymptotic de Sitter spacetimes, we get
\begin{widetext}
\br
\begin{array}{|cc|}
e^{5 i \alpha_+} - e^{\frac{2 \pi \omega}{\kappa_h}} e^{i \alpha_+} &  
e^{5 i \alpha_-} - e^{\frac{2 \pi \omega}{\kappa_h}} e^{i \alpha_-} \\
e^{7 i \alpha_+} - e^{-\frac{2 \pi \omega}{\kappa_c}} e^{-i \alpha_+} & 
e^{7 i \alpha_-} - e^{-\frac{2 \pi \omega}{\kappa_c}} e^{-i \alpha_-}
\end{array}
= 0 
& \Rightarrow & 
\begin{array}{|cc|}
e^{-\frac{\pi \omega}{\kappa_h}} e^{2 i \alpha_+} - 
e^{\frac{\pi \omega}{\kappa_h}} e^{-2 i \alpha_+} &  
e^{-\frac{\pi \omega}{\kappa_h}} e^{2 i \alpha_-} - 
e^{\frac{\pi \omega}{\kappa_h}} e^{-2 i \alpha_-} \\
e^{\frac{\pi \omega}{\kappa_c}} e^{4 i \alpha_+} - 
e^{-\frac{\pi \omega}{\kappa_c}} e^{-4 i \alpha_+} &  
e^{\frac{\pi \omega}{\kappa_c}} e^{4 i \alpha_-} - 
e^{-\frac{\pi \omega}{\kappa_c}} e^{-4 i \alpha_-} \\
\end{array}
=0 \, ,
\label{eq:detds1}
\er
\end{widetext}
which leads to
\be
\label{eq:detds2}
\begin{array}{|cc|}
\sinh \left( \frac{\pi \omega}{\kappa_h} - \frac{i \pi}{2} \left( 1+j \right) \right) &
\sinh \left( \frac{\pi \omega}{\kappa_h} - \frac{i \pi}{2} \left( 1-j \right) \right) \\
\sinh \left( \frac{\pi \omega}{\kappa_c} + i \pi \left( 1+j \right) \right) &
\sinh \left( \frac{\pi \omega}{\kappa_c} + i \pi \left( 1-j \right) \right) \\
\end{array}
= 0 \, .
\ee
Using properties of hyperbolic functions, we obtain the master
equation for the asymptotic de Sitter spacetimes
(\ref{eq:finres-ds}).

Eliminating $A_{\pm}$ in Eq. (\ref{eq:constraint-af}), for the
asymptotic flat spacetimes, we get
%
\ba
\label{eq:detaf1}
\begin{array}{|cc|}
e^{5 i \alpha_+} - e^{\frac{2 \pi \omega}{\kappa_h}} e^{i \alpha_+} &  
e^{5 i \alpha_-} - e^{\frac{2 \pi \omega}{\kappa_h}} e^{i \alpha_-} \\
e^{-i \alpha_+} & 
e^{-i \alpha_-}
\end{array}
&=& 0 \, ,  
\ea
\ba
\label{eq:detaf2}
\begin{array}{|cc|}
e^{5 i \alpha_+} - e^{\frac{2 \pi \omega}{\kappa_h}} e^{i \alpha_+} &  
e^{5 i \alpha_-} - e^{\frac{2 \pi \omega}{\kappa_h}} e^{i \alpha_-} \\
e^{7i \alpha_+} & 
e^{7i \alpha_-}
\end{array}
&=& 0 \,  
\ea
for the two cases. 
Simplifying the above expression, we obtain the master 
equation for the asymptotic flat spacetimes
(\ref{eq:finres-afs}).

\subsection{Asymptotic anti-de Sitter spacetimes}

Eliminating $A_+$, $A_-$ from
(\ref{eq:constraint1ads},\ref{eq:constraint2ads}), we get 
\begin{widetext}
{\small
\ba
\label{eq:detads1}
\begin{array}{|cc|}
e^{i (\alpha_+ + \beta_+ -\omega x_0) } - 
e^{-i(\alpha_+ + \beta_+ - \omega x_0)} & 
e^{i (\alpha_- + \beta_+ - \omega x_0)} - 
e^{-i (\alpha_- + \beta_+ - \omega x_0)} \\
e^{-3 i \alpha_+} & e^{-3 i \alpha_-}
\end{array}
= 0 & \Longrightarrow & 
\begin{array}{|cc|}
\sin \left( \alpha_+ + \beta_+ - \omega x_0 \right) & 
\sin \left( \alpha_- + \beta_+ - \omega x_0 \right) \\
e^{-3 i \alpha_+} &
e^{-3 i \alpha_-}
\end{array}
= 0  \, . \nn 
\ea
}
\end{widetext}
Simplifying, we get,
\beq
\omega x_0 = \frac{\pi}{4} + \beta_+ - i \tanh^{-1} 
\left( \frac{ \tan \left( \frac{\pi}{4}j \right)}{\tan 
\left( \frac{3 \pi}{4}j \right)} \right) 
\eeq
Using properties of hyperbolic functions, we get the master equation
for the asymptotic anti-de Sitter spacetimes (\ref{eq:resultads}).



\end{document}